\documentclass[prb,twocolumn,showpacs,preprintnumbers,amsmath,amssymb,aps,10pt]{revtex4-1}

\usepackage{graphicx}
\usepackage{subfigure}
\usepackage{bm}
\usepackage[final=true]{hyperref}

\newcommand{\m}{\mathbf}

\begin{document}

\title{Thermodynamically stable skyrmion lattice in tetragonal frustrated antiferromagnet with dipolar interaction}

\author{Oleg I. Utesov$^{1,2}$}
\email{utiosov@gmail.com}

\affiliation{$^1$National Research Center ``Kurchatov Institute'' B.P.\ Konstantinov Petersburg Nuclear Physics Institute, Gatchina 188300, Russia}
\affiliation{$^2$Department of Physics, Saint Petersburg State University, 198504 St.Petersburg, Russia}

\begin{abstract}

Motivated by recent experimental results on GdRu$_2$Si$_2$ [Khanh, N.D., Nakajima, T., Yu, X. et al., \emph{Nat. Nanotechnol.} \textbf{15}, 444-449 (2020)], where nanometric square skyrmion lattice was observed, we propose simple analytical mean-field description of the high-temperature part of the phase diagram of centrosymmetric tetragonal frustrated antiferromagnets with dipolar interaction in the external magnetic field. In the reciprocal space dipolar forces provide momentum dependent biaxial anisotropy. It is shown that in tetragonal lattice in the large part of the Brillouin zone for mutually perpendicular modulation vectors in the $ab$ plane this anisotropy has mutually perpendicular easy axes and collinear middle axes, what leads to double-Q modulated spin structure stabilization. The latter turns out to be a square skyrmion lattice in the large part of its stability region with the topological charge $\pm 1$ per magnetic unit cell, which is determined by the frustrated exchange coupling, and, thus, nanometer-sized. In the presence of additional single-ion easy-axis anisotropy, easy and middle axes can be swapped, which leads to different phase diagram. It is argued that the latter case is relevant to GdRu$_2$Si$_2$.

\end{abstract}
\maketitle

\section{Introduction}
\label{Sintro}

Originally, skyrmions were proposed by T. Skyrme in 1962 in order to describe nucleons as topologically stable field configurations~\cite{skyrme1962}. In magnetism skyrmions first emerge as metastable states in two-dimensional ferromagnets in Ref.~\cite{belavin1975metastable}. Crucial next steps were made in seminal papers~\cite{bogdanov1989,bogdanov1994} where it was shown that single skyrmions and skyrmion lattices (SkL) can be stabilized in noncentrosymmetric magnets due to the Dzyaloshinskii-Moriya interaction~\cite{dzyaloshinsky1958,moriya1960} (DMI). Finally, after experimental observation of the SkL in MnSi in the so-called A phase~\cite{muhlbauer2009}, magnetic skyrmions become one of the hottest topics of the contemporary physics (see, e.g., Refs.~\cite{fert2017,bogdanov2020} for review). Importantly, this interest is stimulated by promising technological applications, one of which is the racetrack memory~\cite{fert2013}.

Efficiency of the possible nanodevices relies on magnetic skyrmions non-trivial topology~\cite{belavin1975metastable}. Topological charge of magnetic structure is defined as the spin direction winding number on a unit sphere,
\begin{equation}\label{charge1}
  Q = \frac{1}{4 \pi} \int \m{n} \cdot \left[ \partial_x \m{n} \times \partial_y \m{n} \right] dx dy,
\end{equation}
where $\m{n} = \m{s}/|\m{s}|$ is a unit vector along the averaged over thermodynamical (and/or quantum) fluctuations spin direction. For individual skyrmion integral over its size usually yields $Q = \pm 1$, whereas for the SkL natural measure is a density of topological charge, $n_{sk}$. The latter quantity is of prime importance as, for instance, topological contribution to Hall resistivity $\rho^T \propto n_{sk}$~\cite{neubauer2009}. Note that other non-trivial magnetic textures are actively studied, see Ref.~\cite{gobel2020} for review.

It was understood recently that skyrmions can be stabilized not only in systems with DMI but also in frustrated centrosymmetric systems~\cite{leonov2015} due to anisotropic interactions. This effect was indeed observed in centrosymmetric frustrated triangular-lattice magnet Gd$_2$PdSi$_3$~\cite{kurumaji2019SkL}. Importantly, frustration is crucial in many multiferroics of spin origin~\cite{nagaosa}, and skyrmions can lead to interesting effects in such materials~\cite{kurumaji2019Rev}.

Recent observation of the SkL in the centrosymmetric tetragonal material GdRu$_2$Si$_2$~\cite{khanh2020} stimulates related theoretical researches~\cite{hayami2020square,wang2020meron}. In these papers low-temperature part of the phase diagram was considered and various phases (including topologically non-trivial) were shown to emerge depending on the anisotropy parameters and the external magnetic field.

In the present study we propose dipolar forces as the stabilizing mechanism of nanometer-sized skyrmions in tetragonal frustrated antiferromagnets. Previously, to the best of our knowledge, in the context of skyrmions magnetic dipolar interaction was only considered as leading to large micrometer-sized magnetic bubbles~\cite{bogdanov2020,hubert1998}. Moreover, our analytical mean-field (Landau) approach is unusually simple in the context of topologically non-trivial spin textures.

Dipolar interaction is often small and, thus, negligible. However, in some materials, e.g., RbFeCl$_3$~\cite{shiba,gekht}, MnBr$_2$~\cite{mnbr2}, MnI$_2$~\cite{Utesov2017}, it was shown to be important anisotropic coupling. From the general arguments it should be correct for materials with magnetic ions in spherically-symmetrical state with $L=0$, because other anisotropic interactions are moderated by the spin-orbit coupling~\cite{white}. Furthermore, dipolar forces can lead to rather complicated sequences of phase transition at large temperatures~\cite{shiba,Utesov2017} and small temperatures in magnetic field~\cite{utesov2019,utesov2020phase}. Note, that in GdRu$_2$Si$_2$ magnetic Gd$^{3+}$ ions~\cite{slaski1984} are in state with $S = 7/2$ and $L =0$.

\begin{figure}
  \centering
  \hfill
  \includegraphics[width=4cm]{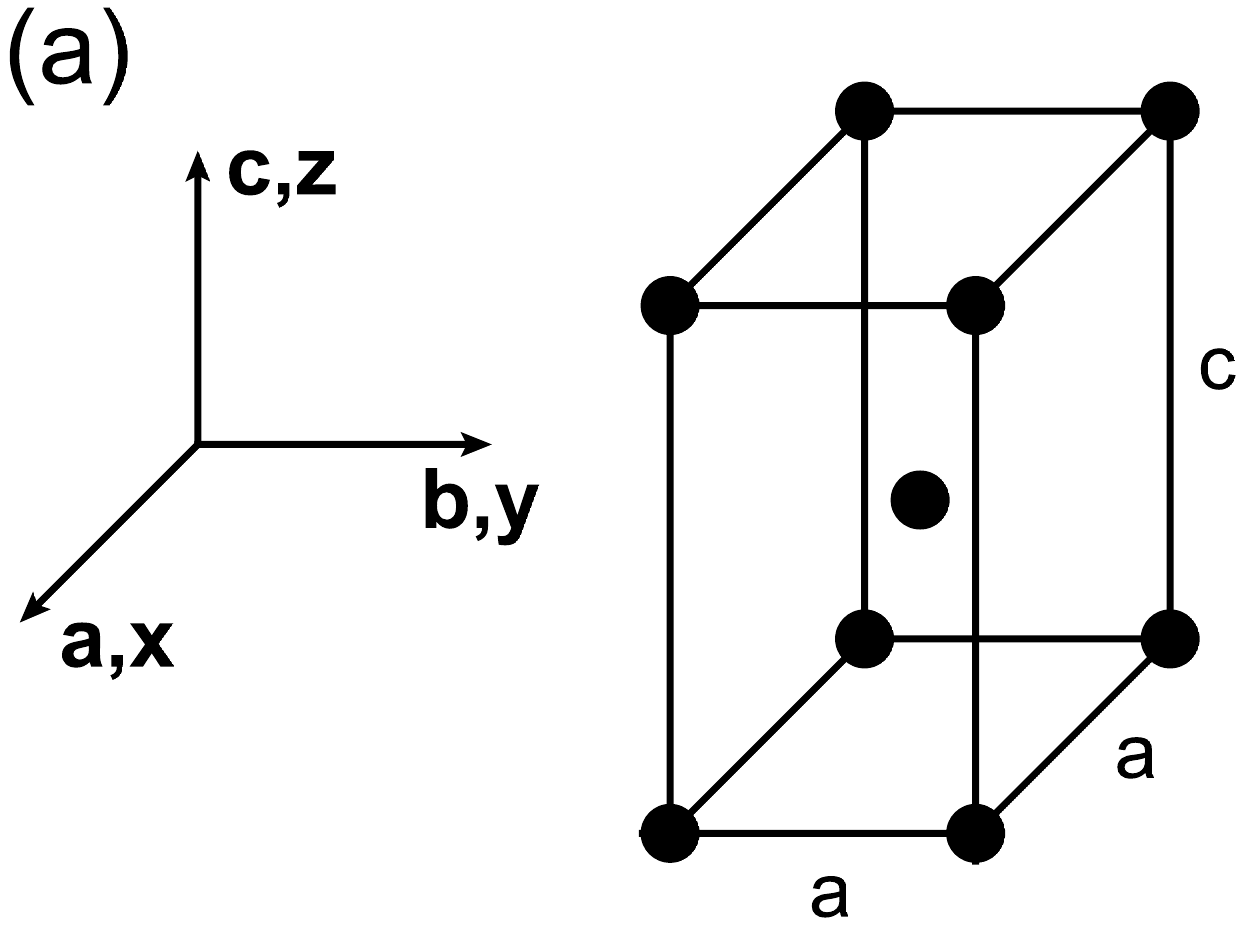}
  \hfill
  \includegraphics[width=4cm]{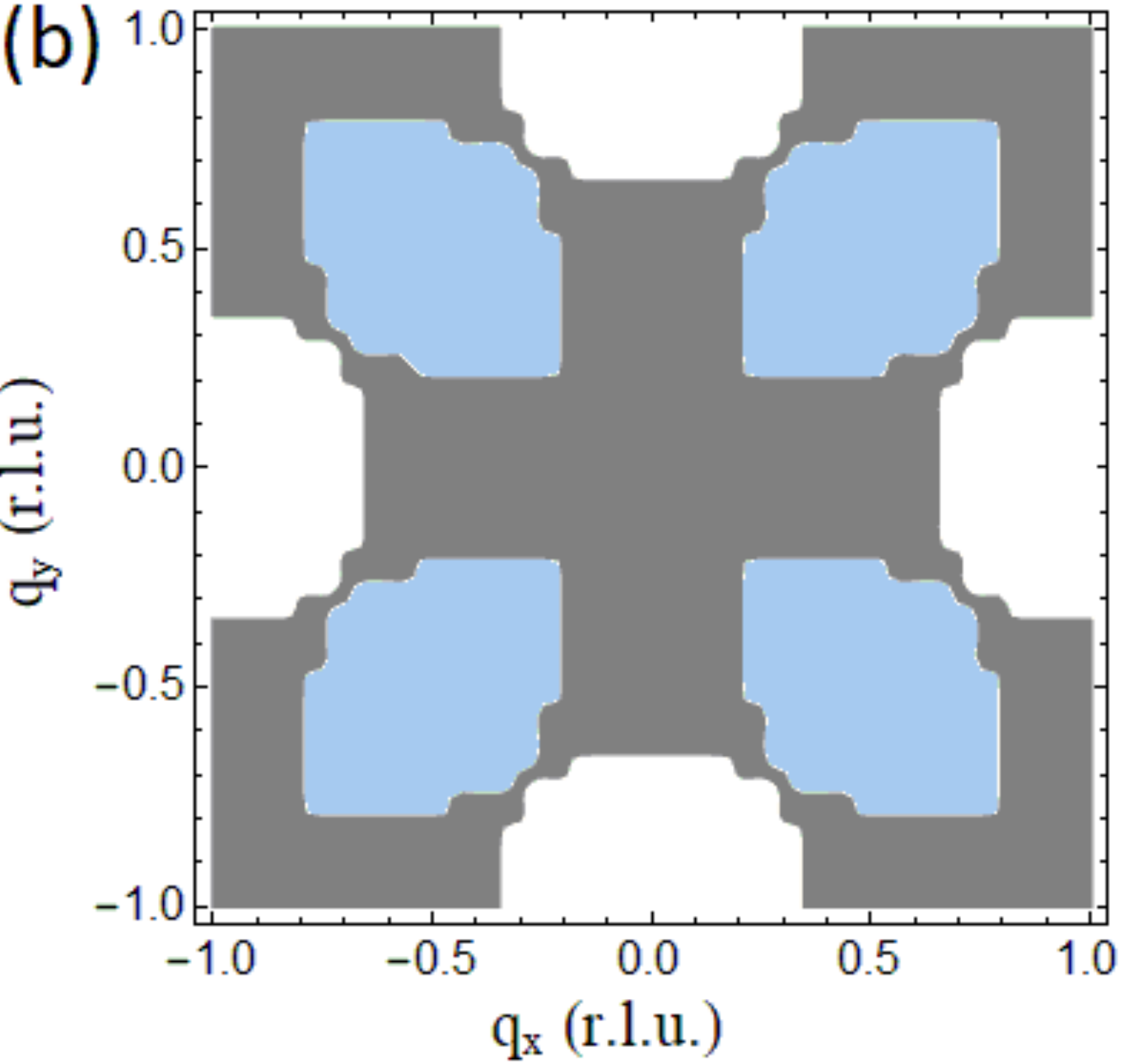}
  \hfill
  \vspace{1cm}
  \centering
  \includegraphics[width=3cm]{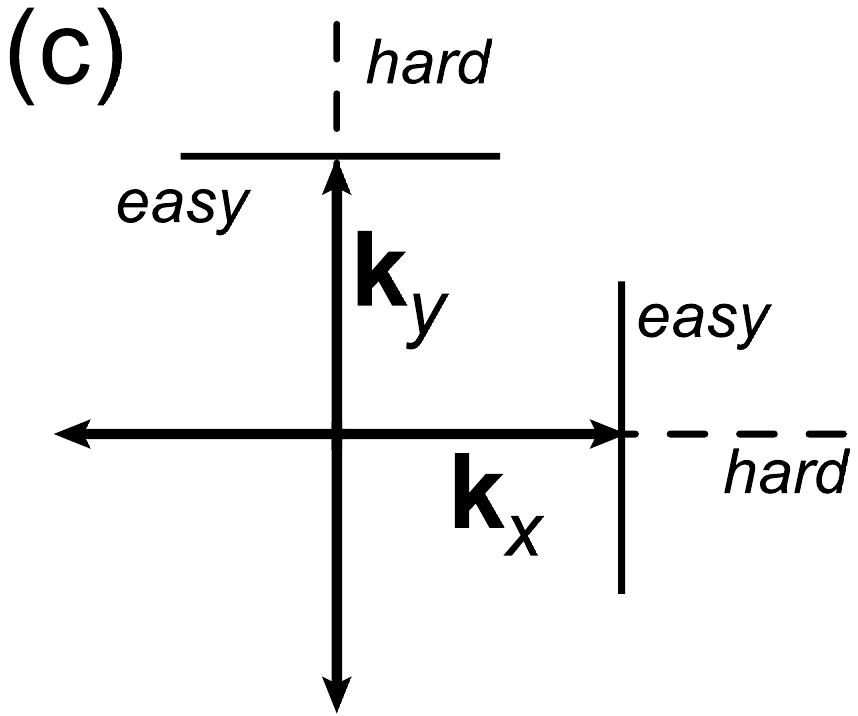}
  \caption{(a) Relevant to the present study tetragonal structure of GdRu$_2$Si$_2$; only magnetic Gd$^{3+}$ ions are shown. (b) Using lattice parameters $a = 4.165$ \AA \, and $c = 9.61$ \AA \, of Ref.~\cite{slaski1984} one can calculate Fourier transform of the dipolar tensor (see Eq.~\eqref{dip2}) numerically for $|\m{q}| \neq 0$. Not taking into account other possible anisotropic interactions, we found that for in-plane modulation vectors $\mathbf{q}=(q_x,q_y,0)$  the $\m{c}$ axis is the easy, middle, and hard one in blue, gray, and white region of the Brillouin zone, correspondingly. (c) For vectors $\pm \mathbf{k}_x = (k,0,0)$ and $\pm \mathbf{k}_y=(0,k,0)$, in a wide range of $k$, easy axes are mutually perpendicular and middle ones are collinear, oriented along $\m{c}$ (additional single-ion anisotropy can swap the easy and middle axes). This property of the dipolar interaction, which sometimes is referred to as the \emph{compass anisotropy}~\cite{banerjee2013,chen2016exotic,wang2020meron}, leads to square skyrmion lattice stabilization in a certain part of the phase diagram.}\label{Fig1}
\end{figure}

Our model is based on a simple property of dipolar forces in tetragonal magnets which provide effective momentum-dependent biaxial anisotropy. In particular case when the modulation vector $\m{q}$ lies in the $ab$ plane (conventional basis vectors $\m{a} \perp \m{b} \perp \m{c}, \, |\m{a}|=|\m{b}|$ are used) in the large part of the Brillouin zone (BZ) the easy axis lies in-plane and the middle one is along $\m{c}$, or \emph{vice versa} (see Fig.~\ref{Fig1}). This leads to energetically effective combining of elliptical spirals with mutually perpendicular in-plane modulation vectors into double-Q structures.

\begin{figure}
  \centering
  \includegraphics[width=4cm]{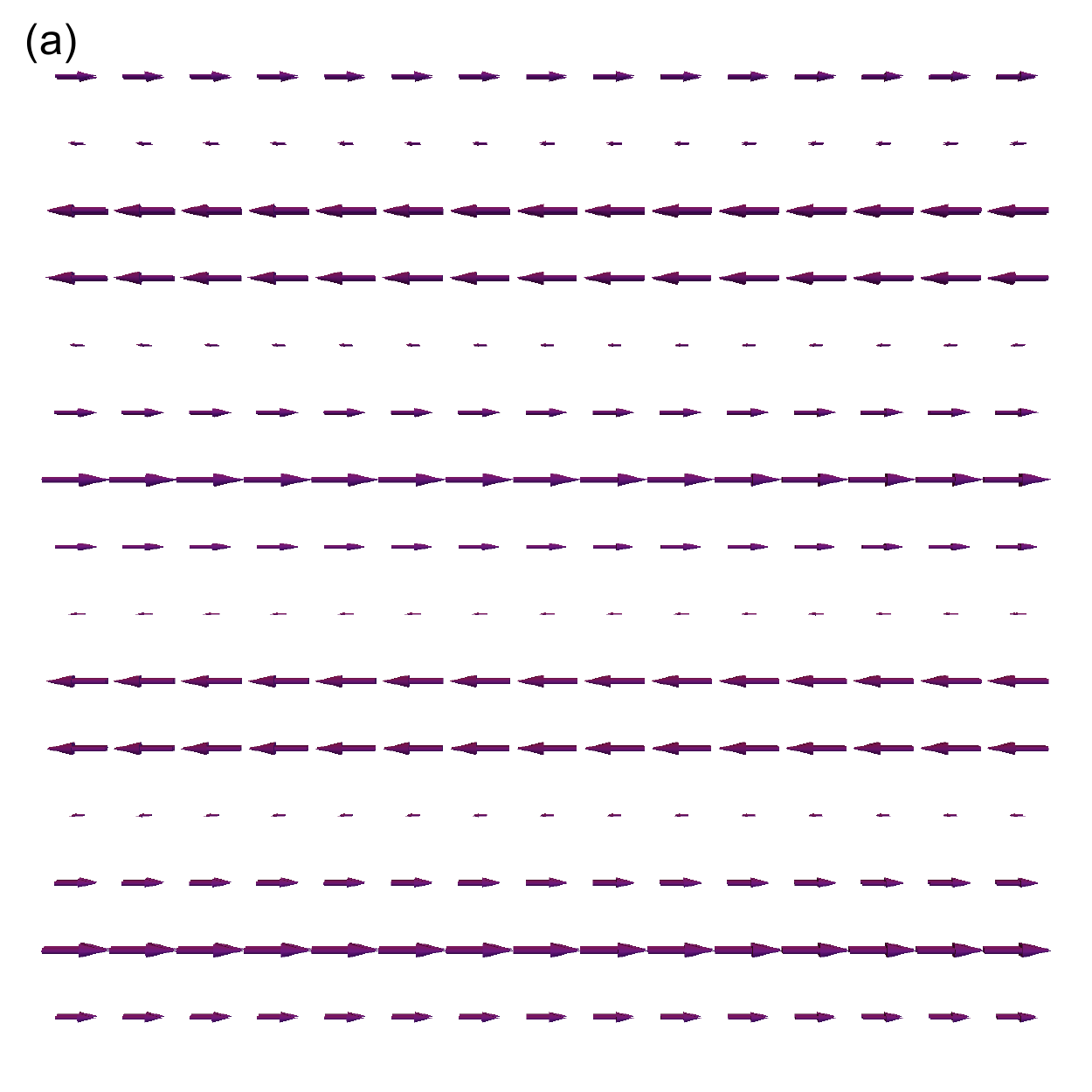}
  \includegraphics[width=4cm]{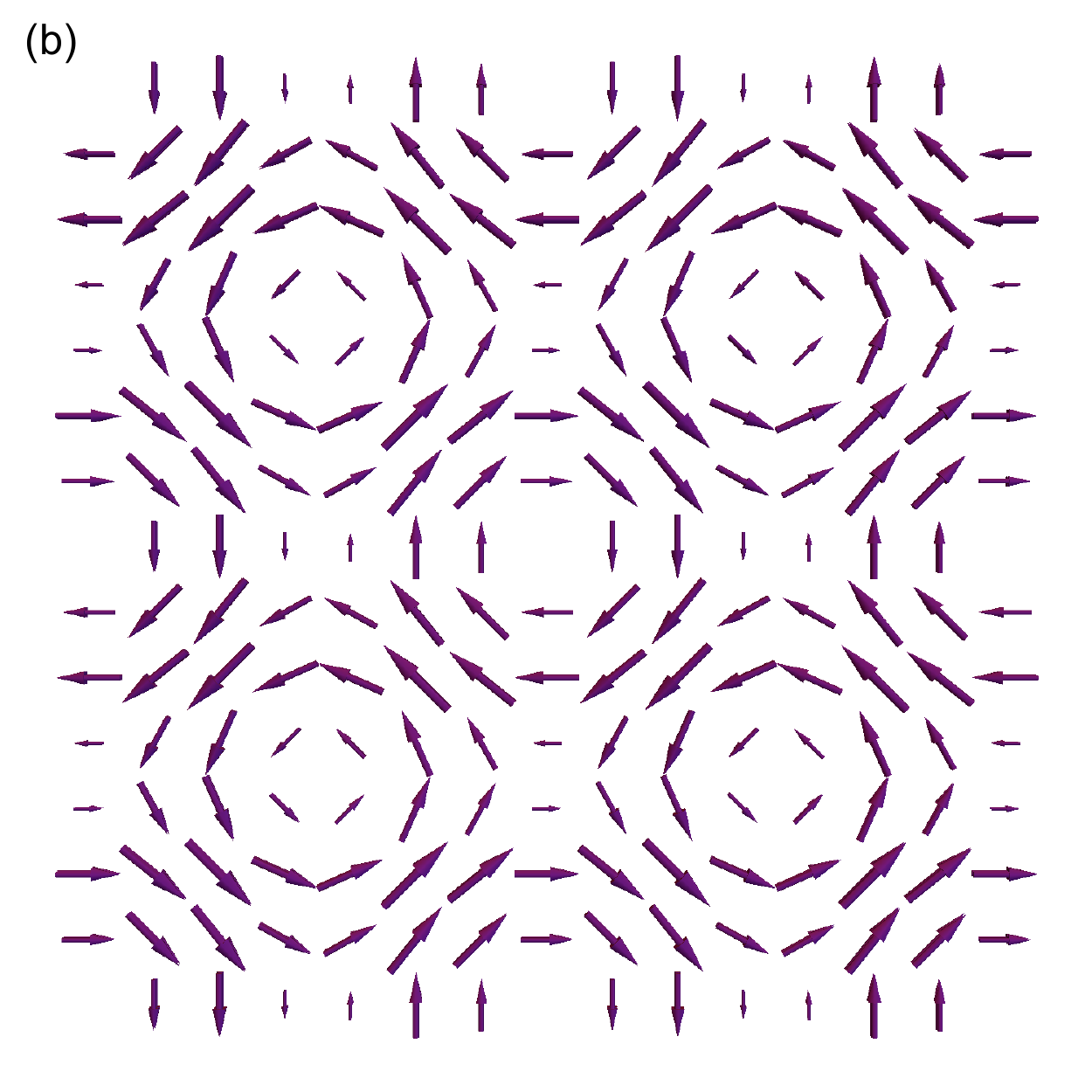}
  \centering
  \includegraphics[width=4cm]{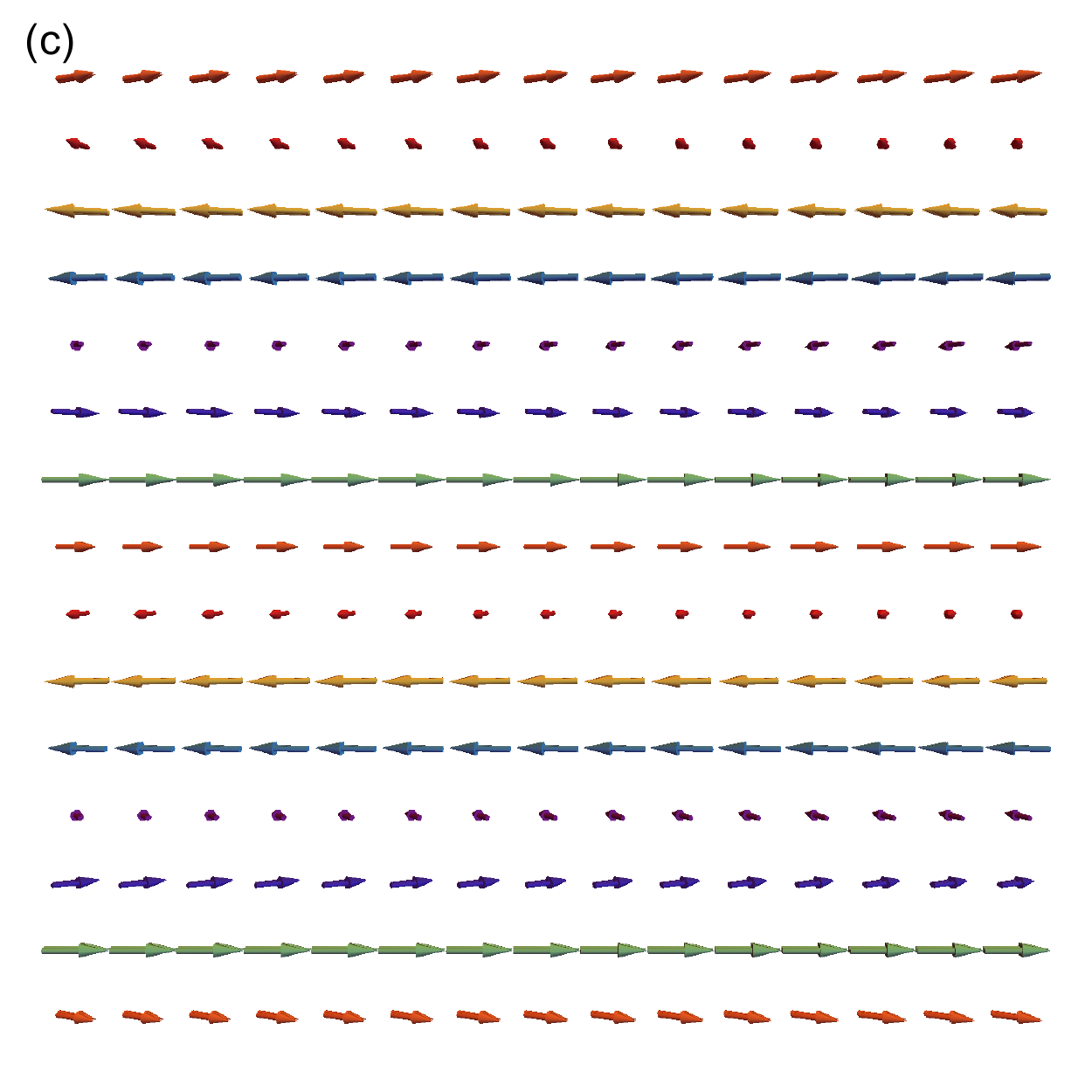}
  \includegraphics[width=4cm]{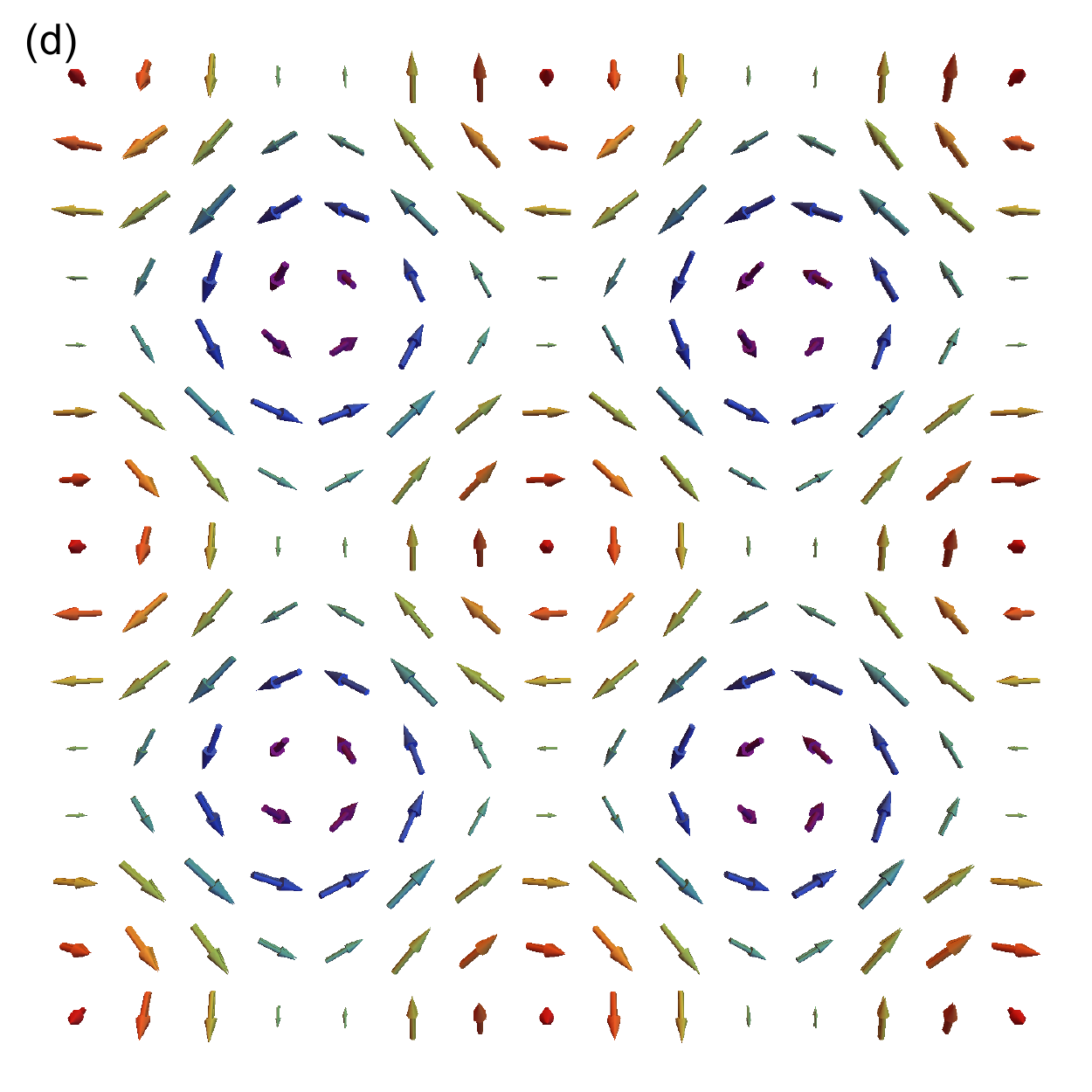}
  \caption{Possible magnetic structures at zero external field in the $ab$ plane; region of $2\times2$ cells (with size $2 \pi/k$) is shown. (a) Sinusoidal single-modulated spin-density wave (1S). (b) Vortical double spin-density wave (2S). (c) Elliptical spiral (1Q). (d) Double-Q elliptical phase which consists of alternating merons and anti-merons (2Q). In the first two structures spins lie in-plane. External field uniformly magnetize them and transform into simple and double fan structures, respectively. For the latter two orderings $z$-component of spins is shown by rainbow colors (from red -- spin-up state to magenta -- spin-down state).}\label{FigSpins}
\end{figure}

Using mean-field approach we show that in relevant to experimental results of Ref.~\cite{khanh2020} case of two possible modulation vectors along $\m{a}$ and $\m{b}$ axes (without additional single-ion anisotropy) peculiar sequence of phase transitions is realized in large temperatures domain of the phase diagram. First, upon temperature lowering the system undergoes second order phase transition from paramagnetic phase to vortical double spin-density wave state, which will be referred to as 2S, see Fig.~\ref{FigSpins}(b). Next, components of the order parameters along the middle axis emerge which manifests continuous transition from 2S to the spin structure with two elliptical screw spirals combined (2Q, see Fig.~\ref{FigSpins}(d)). Finally, there is a first order phase transition from the 2Q to single-Q elliptical spiral (1Q, see Fig.~\ref{FigSpins}(c)). Importantly, at nonzero magnetic fields along the $\m{c}$ axis, part of the phase diagram region where the 2Q structure is the ground state becomes topologically nontrivial, being a square SkL with one (anti)skyrmion per magnetic unit cell. We also show that if the single-ion easy-axis anisotropy (which allows to swap easy and middle axes) is added into consideration the phase diagram can drastically change, and the square SkL emerges only at magnetic fields exceeding a certain finite value. In this case our approach qualitatively reproduces experimentally observed phase diagram of GdRu$_2$Si$_2$~\cite{khanh2020}.

The rest of the paper is organized as follows. In Sec.~\ref{Smodel} we introduce the spin Hamiltonian which consists of frustrated exchange coupling, dipolar interaction, and the Zeeman term. We also formulate the mean-field approach and discuss relevant parameters. Section~\ref{SMF} is devoted to the mean-field analysis of the high temperature part of the temperature-magnetic field phase diagram for the case of mutually perpendicular easy axes. Free energies of the relevant spin structures are derived, and the phase boundaries are determined. In Sec.~\ref{Stop} we discuss topological properties of the 2Q phase and show that in a certain part of the corresponding region of the phase diagram it is a square SkL.  Section~\ref{SOOP} addresses the case of collinear easy axes and mutually perpendicular middle ones, and a relevance to experimental findings of Ref.~\cite{khanh2020}. Finally, Sec.~\ref{Sdis} summarizes our results and contains related discussion.

\section{Model}
\label{Smodel}

We consider frustrated antiferromagnet on a tetragonal lattice (both simple and body-centered) with one magnetic ion in a unit cell. System Hamiltonian also includes magneto-dipolar interaction, single-ion anisotropy, and Zeeman term, being
\begin{eqnarray}
 \label{ham1}
  \mathcal{H} &=& \mathcal{H}_{ex} + \mathcal{H}_{d} + \mathcal{H}_s + \mathcal{H}_{z}, \nonumber \\
  \mathcal{H}_{ex} &=& -\frac12 \sum_{i,j} J_{ij} \left(\mathbf{S}_i \cdot \mathbf{S}_j\right), \nonumber \\
  \mathcal{H}_d &=& \frac12 \sum_{i,j} D^{\alpha \beta}_{ij} S^\alpha_i S^\beta_j,  \\
  \mathcal{H}_s &=& - Z \sum_{i} (S^z_i)^2, \nonumber \\
  \mathcal{H}_z &=& - \sum_i \left(\mathbf{h} \cdot \mathbf{S}_i\right).\nonumber
\end{eqnarray}
Here $\mathbf{h}= g \mu_B \mathbf{H}$ is the external magnetic field in energy units, $\alpha, \beta$ denotes cartesian coordinates. For spin components we use conventional global basis with $z$ coordinate along the $\m{c}$ axis, $x$ and $y$ along edges of the unit cell in the $ab$-plane (see Fig.~\ref{Fig1}). Dipolar tensor is given by
\begin{equation}\label{dip1}
	 {\cal D}^{\alpha \beta}_{ij} = \omega_0 \frac{v_0}{4 \pi} \left( \frac{1}{R_{ij}^3} - \frac{3 R_{ij}^\alpha R_{ij}^\beta }{R_{ij}^5}\right),
\end{equation}
where $v_0$ is a unit cell volume. Characteristic energy of the dipole interaction reads
\begin{equation}\label{dipen}
  \omega_0 = 4 \pi \frac{(g \mu_B)^2}{v_0}.
\end{equation}
This anisotropic interaction is of prime importance for magnetic ions with half-filled electronic shell, e.g., Mn$^{2+}$ or Eu$^{2+}$. For such ions $L=0$ and dipolar forces are usually one of the most important anisotropic terms.

After Fourier transform ($N$ is a total number of spins)
\begin{equation}
\label{four1}
  \mathbf{S}_j = \frac{1}{\sqrt{N}} \sum_\mathbf{q} \mathbf{S}_\mathbf{q} e^{i \mathbf{q} \mathbf{R}_j},
\end{equation}
Hamiltonian~\eqref{ham1} acquires the following form:
\begin{eqnarray}
  \label{ex2}
  \mathcal{H}_{ex} &=& -\frac12 \sum_\mathbf{q} J_\mathbf{q} \left(\mathbf{S}_\mathbf{q} \cdot \mathbf{S}_{-\mathbf{q}}\right), \\
	\label{dip2}
  \mathcal{H}_d &=& \frac12 \sum_\mathbf{q} {\cal D}^{\alpha \beta}_\mathbf{q} S^\alpha_\mathbf{q} S^\beta_{-\mathbf{q}}. \\
	\label{sinion1}
\mathcal{H}_s &=& Z \sum_\mathbf{q}  S^z_\mathbf{q} S^z_{-\mathbf{q}}. \\
	\label{z21}
 \mathcal{H}_z &=& - \sqrt{N} \left(\mathbf{h} \cdot \mathbf{S}_{\bf 0}\right).
\end{eqnarray}
Importantly, the first three terms here can be combined into
\begin{equation}\label{tens1}
  \mathcal{H}_0 = - \sum_\mathbf{q} \mathcal{H}^{\alpha\beta}_\mathbf{q} S^\alpha_\mathbf{q} S^\beta_{-\mathbf{q}},
\end{equation}
where ``$0$'' denotes the Hamiltonian at $h=0$. Tensor $\mathcal{H}^{\alpha\beta}_\mathbf{q}$ has three eigenvalues $\lambda_1(\mathbf{q}) \geq \lambda_2(\mathbf{q}) \geq \lambda_3(\mathbf{q})$ corresponding to three eigenvectors $\mathbf{v}_1(\mathbf{q}), \, \mathbf{v}_2(\mathbf{q}), \, \mathbf{v}_3(\mathbf{q})$ at each momentum. The latter define particular basis of easy, middle, and hard axes for each $\mathbf{q}$. This momentum-dependent biaxial anisotropy is due to dipolar forces.

Dipolar tensor in the reciprocal space ${\cal D}^{\alpha \beta}_\mathbf{q}$ can be calculated numerically using standard technique involving rewriting it in a fast convergent form (see Ref.~\cite{cohen} and references therein). Moreover, at large temperatures (close to the transition to the paramagnetic phase) only particular $\mathbf{q}$ are important which significantly simplifies corresponding analysis~\cite{Utesov2017}. Since dipolar forces are usually small in comparison with exchange coupling, these momenta are close to those where $J_\mathbf{q}$ has (local) maxima, which are assumed to be incommensurate due to frustration. Thus, at small temperatures and small $h$ some sort of a spiral ordering is the ground state of the system.

Below we shall mostly discuss particular case where magnetic ordering modulation vectors are oriented along $\m{a}$ and $\m{b}$ axes, being $\mathbf{k}_x = (k,0,0)$ and $\mathbf{k}_y=(0,k,0)$. Not taking into account possible effect of the single-ion anisotropy we arrive to crucial point for the present theory: in a wide range of parameters of tetragonal lattice it can be shown numerically that the easy axis for $\mathbf{k}_x$ is $\m{b}$, the hard one is $\m{a}$, and \emph{vice versa} for $\mathbf{k}_y$. The middle axis is $\m{c}$ for the both vectors (see Fig.~\ref{Fig1}). This exactly realizes in case of GdRu$_2$Si$_2$, where $k=0.22$ in the reciprocal lattice units~\cite{khanh2020} (in used below notation $k=0.22 \times 2 \pi/a$). Furthermore, this provides a simple physical ground for anisotropic momentum-dependent terms used in recent theoretical studies~\cite{hayami2020square,wang2020meron}, the compass anisotropy which was previously attributed to the spin-orbit coupling~\cite{banerjee2013}.

We point out that the frustration can lead to competition with incommensurate structures characterized by another momenta with close value of $J_\mathbf{q}$. In general case, corresponding local axes basis will not possess the feature described above. For instance, if the modulation vector $\m{q} || \m{c}$ dipolar tensor simply makes $ab$ plane an easy one. It can further complicate the phase diagram introducing some additional intermediate phases.

In our high-temperature calculations we shall use $\mathbf{s}_i$ for mean value of the corresponding spin operator $\m{S}_i$. It can be shown that the free energy can be expressed as (see, e.g., Refs.~\cite{gekht1984, Utesov2017} for details)
\begin{equation}\label{Free1}
  \mathcal{F} = - \sum_\mathbf{q} \mathcal{H}^{\alpha\beta}_\mathbf{q} s^\alpha_\mathbf{q} s^\beta_{-\mathbf{q}} - \sqrt{N} \mathbf{h} \cdot \mathbf{s}_{\bf 0} + A T \sum_i s^2_i + B T_c \sum_i s^4_i,
\end{equation}
provided that $|\m{s}_i| \ll S$; $T_c=\lambda_1(\mathbf{k}_x)/A$ is the temperature of the phase transition from paramagnetic to magnetically ordered phase at $h=0$, which will be specified later. Expansion parameters $A$ and $B$ are given by
\begin{eqnarray}
  A &=& \frac{3}{2S(S+1)}, \\
  B &=& \frac{9[(2S+1)^4-1]}{20 (2S)^4(S+1)^4}.
\end{eqnarray}
For $S=7/2$ one has $A \approx 0.095$ and $B \approx 0.002$.

In order to make a connection with real materials we estimate relevant parameters using experimental data of Ref.~\cite{khanh2020}. For the case without single-ion anisotropy ($Z=0$) using only ordering temperature $T_c(B=0) \approx 45$ K, saturation field $B_{sat}(T=0) \approx 10$ T (in energy units $h_{sat} \approx S(J_{\m{k}_x}-J_\m{0})$ if one neglects small anisotropy and shape-dependent corrections, see e.g. Ref.~\cite{utesov2020phase}), and numerically calculated dipolar tensor we get (all values are in Kelvins)
\begin{eqnarray} \label{Par1}
  \lambda_1 &\approx& 4.3, \quad J_\m{0} \approx 4.6, \\
  \lambda_1 - \lambda_2 &\approx& 0.05, \quad \lambda_1 - \lambda_3 \approx 0.20,  \nonumber
\end{eqnarray}
where $\lambda_i$ are the same for $\m{k}_x$ and $\m{k}_y$ momenta.

\section{Mean-field approach for in-plane easy axes}
\label{SMF}

In this section we perform mean-field analysis basing on the order parameters smallness at high temperatures. For definiteness, we consider particular case of possible modulation vectors and corresponding axes sets depicted in Fig.~\ref{Fig1}(c).

\subsection{Spin structures at $h=0$}
\label{SSh0}

We start from the simplest case without the external field. In the systems with tetragonal symmetry due to four energy minima at $\pm \mathbf{k}_x$ and $\pm \mathbf{k}_y$, along with conventional single-modulated sinusoidal spin-density wave (SDW) and elliptical (helicoidal) phases, double structures can emerge. Below we calculate free energy for each of relevant spin structures, shown in Fig.~\ref{FigSpins}.

\subsubsection{Single-Q spin-density wave (1S)}

In this case (taking for definiteness $\mathbf{k}_x$ as a modulation vector, $\mathbf{k}_y$ evidently yields the same result)
\begin{equation}\label{Ssdw1}
  \mathbf{s}_i = s \mathbf{e}_y \cos{\mathbf{k}_x \mathbf{R}_i}.
\end{equation}
Using Eq.~\eqref{Free1} we get
\begin{equation}\label{Fsdw1}
  \frac{\mathcal{F}}{N} = - \frac{s^2 \lambda_1}{2} + \frac{s^2 A T}{2} + \frac{3 s^4 B T_c}{8}.
\end{equation}
Minimization with respect to $s$ gives (for $T \leq T_c$)
\begin{equation}\label{Ssdw2}
  s^2 = \frac{2(\lambda_1-AT)}{3BT_c},
\end{equation}
and
\begin{equation}\label{Fsdw2}
  \frac{\mathcal{F}_{1S}}{N} = - \frac{(\lambda_1-AT)^2}{6BT_c}.
\end{equation}

\subsubsection{Double-Q spin-density wave (2S)}

According to the symmetry of the system double-Q spin-density wave structure with both order parameters along the local easy axis becomes possible. Corresponding spin ordering reads
\begin{equation}\label{Ssdw3}
  \mathbf{s}_i = s (\mathbf{e}_y \cos{\mathbf{k}_x \mathbf{R}_i} + \mathbf{e}_x \cos{\mathbf{k}_y  \mathbf{R}_i}).
\end{equation}
In real space this is a vortex structure depicted in Fig.~\ref{FigSpins}(b). Note that the phases of trigonometric functions are not important here and can be taken arbitrary due to the system translational invariance and incommensurability of the modulation vector.
Using Eq.~\eqref{Free1} one gets
\begin{equation}\label{Fsdw3}
  \frac{\mathcal{F}}{N} = - s^2 \lambda_1 + s^2 A T + \frac{5 s^4 B T_c}{4}.
\end{equation}
Minimization with respect to $s$ yields
\begin{equation}\label{Ssdw4}
  s^2 = \frac{2(\lambda_1-AT)}{5BT_c},
\end{equation}
and
\begin{equation}\label{Fsdw4}
  \frac{\mathcal{F}_{2S}}{N} = - \frac{(\lambda_1-AT)^2}{5BT_c}.
\end{equation}
The last quantity is always smaller than the free energy of the single-Q SDW~\eqref{Fsdw2}. As a corollary, at $T_c$ system undergoes phase transition between paramagnetic phase and double-Q vortex structure. Complementary low temperature result at high magnetic field along the $\m{c}$ axis is the appearance of magnetized along the field and vortical in perpendicular plane double-Q phase instead of single-Q fan one~\cite{hayami2020square}.

\subsubsection{Single-Q elliptical phase (1Q)}

We further proceed with modulated along one direction ($\m{k}_x$ is taken for definiteness) elliptical structure:
\begin{equation}\label{Sel1}
  \mathbf{s}_i = s_1 \mathbf{e}_y \cos{\mathbf{k}_x \mathbf{R}_i} + s_2 \mathbf{e}_z \sin{\mathbf{k}_x  \mathbf{R}_i}.
\end{equation}
Chirality of this structure is not important; one can freely vary the sign of the second term and the common for sine and cosine functions phase.

Corresponding free energy reads
\begin{eqnarray}\label{Fel1}
  \frac{\mathcal{F}}{N} &=& - \frac{s_1^2 \lambda_1 + s_2^2 \lambda_2}{2} + \frac{(s_1^2+s^2_2) A T}{2} \\ &+& \frac{(3 s_1^4 + 2 s^2_1s^2_2 + 3 s^4_2) B T_c}{8} \nonumber.
\end{eqnarray}
Nonzero $s_2$ emerges at $T < T_{1Q}= T_c - 3(\lambda_1-\lambda_2)/2A$; the spin components are given by
\begin{eqnarray}\label{Sel2}
  s^2_1 &=& \frac{2 (\lambda_1 - A T) + (\lambda_1-\lambda_2)}{4 B T_c}, \\ s^2_2 &=& \frac{2 (\lambda_1 - A T) -  3(\lambda_1-\lambda_2)}{4 B T_c} \nonumber.
\end{eqnarray}
So, the free energy has the following form:
\begin{eqnarray}\label{Fel2}
  &&\frac{\mathcal{F}_{1Q}}{N} =
  \\ && - \frac{4(\lambda_1-AT)^2 - 4(\lambda_1-AT)(\lambda_1-\lambda_2) + 3(\lambda_1-\lambda_2)^2}{16BT_c} \nonumber.
\end{eqnarray}

Below we consider magnetic field along the $\m{c}$ axis, so similar to the 1Q phase can emerge -- the conical phase with spins rotating in $ab$ plane (we shall refer to it as XY). At zero field its energy is given by Eq.~\eqref{Fel2} with the substitution $\lambda_2 \rightarrow \lambda_3$.

\subsubsection{Double-Q elliptical phase (2Q)}

We turn to a superposition of two single-Q elliptical structures with mutually perpendicular modulation vectors $\mathbf{k}_x$ and $\mathbf{k}_y$. This structure will be referred to as 2Q. Corresponding spin arrangement is given by
\begin{eqnarray}\label{Sel3}
  \mathbf{s}_i &=& s_1 (\mathbf{e}_y \cos{\mathbf{k}_x \mathbf{R}_i} + \mathbf{e}_x \cos{\mathbf{k}_y  \mathbf{R}_i}) \\
  &&+ s_2 \mathbf{e}_z (\sin{\mathbf{k}_x \mathbf{R}_i} + \sin{\mathbf{k}_y  \mathbf{R}_i}) \nonumber.
\end{eqnarray}
Once again chiralities and phases of both components can be arbitrary, they do not affect the free energy.

Corresponding free energy reads
\begin{eqnarray}\label{Fel3}
  \frac{\mathcal{F}}{N} &=& - (s_1^2 \lambda_1 + s_2^2 \lambda_2) + (s_1^2+s^2_2) A T \\ &+& \frac{(5 s_1^4 + 6 s^2_1s^2_2 + 9 s^4_2) B T_c}{4} \nonumber.
\end{eqnarray}
This structure is possible if $T < T_{2Q}= T_c - 5(\lambda_1-\lambda_2)/2A$. The order parameters are following:
\begin{eqnarray}\label{Sel4}
  s^2_1 = \frac{2 (\lambda_1 - A T) + (\lambda_1-\lambda_2)}{6 B T_c}, \\ \, s^2_2 = \frac{2 (\lambda_1 - A T) -  5(\lambda_1-\lambda_2)}{18 B T_c} \nonumber,
\end{eqnarray}
and the free energy has the form:
\begin{eqnarray}\label{Fel4}
  && \frac{\mathcal{F}_{2Q}}{N} = \\ && - \frac{8(\lambda_1-AT)^2 - 4(\lambda_1-AT)(\lambda_1-\lambda_2) + 5(\lambda_1-\lambda_2)^2}{36BT_c} \nonumber.
\end{eqnarray}

To conclude this subsection, we point out that it can be shown that the double XY structure has larger free energy in comparison with the simple one and, consequently, should not be considered.

\subsection{Sequence of phase transitions at $h=0$}
\label{SSpth0}

Presented above analytical equations for free energies of different phases implicitly depend on the corresponding structures modulation vectors through $\lambda_i$ ($i=1,2$) $\m{q}$-dependence. For 2S vortical structure (see Eq.~\eqref{Fsdw4}) it is evident that the modulation vector corresponds to $\lambda_1(\m{q})$ maximal value (such a $\m{q}$ is referred to as $\m{k}_x$ or $\m{k}_y$). However, for other phase it is not completely true due to possibly different behaviour of $\lambda_1(\m{q})$ and $\lambda_2(\m{q})$ these points, which can shift the structure modulation vector (it was indeed observed in Ref.~\cite{khanh2020}). Nevertheless, since isotropic exchange interaction is usually much larger than the dipolar forces, we neglect this small effect below and do not write $\lambda_i$ $\m{q}$-dependence.

For consideration of the phase transitions, we first simplify the notation: let $t = \lambda_1 - A T$ (in the magnetically ordered phases $t>0$), and $\Lambda = \lambda_1 - \lambda_2 >0$. Then, one should compare the following ``free energies'':
\begin{eqnarray} \label{Free2}
  f_{2S} &=& -\frac{t^2}{5}, \, \frac{5 \Lambda}{2} \geq t>0 \nonumber \\
  f_{1Q} &=& -\frac{4 t^2 - 4 \Lambda t + 3 \Lambda^2}{16}, \, t>\frac{3 \Lambda}{2}, \\
  f_{2Q} &=& -\frac{8 t^2 - 4 \Lambda t + 5 \Lambda^2}{36}, \, t>\frac{5 \Lambda}{2}.  \nonumber
\end{eqnarray}
The smallest one at given $t$ indicates the ground state of the system.

Naturally, at $t \gg \Lambda$ the 1Q phase (single-Q elliptical spiral) is the ground state. Possible first order phase transition between 2S and 1Q can be determined from the equation:
\begin{equation}\label{PT1}
  \frac{t^2}{5}=\frac{4 t^2 - 4 \Lambda t + 3 \Lambda^2}{16}.
\end{equation}
Corresponding solutions read
\begin{equation}\label{PT2}
  t = \frac{5 \pm \sqrt{10}}{2} \Lambda \approx 0.9 \Lambda; \, 4.1 \Lambda.
\end{equation}
Evidently, they are non-physical: the one with the ``$+$'' sign is larger than $t_{2Q} = 2.5 \Lambda $ at which 2Q structure emerges and substitutes 2S, and another one with ``$-$'' is smaller than $t_{1Q}=1.5\Lambda$ which is the boundary for 1Q structure (meta)stability. Thus, if one neglects a possibility of different values of $k$ for 1Q and 2Q the following scenario of phase transitions upon temperature variation takes place: \mbox{PM $\leftrightarrow$ 2S $\leftrightarrow$ 2Q $\leftrightarrow$ 1Q}. The first two are second order phase transitions. The latter is of the first order; corresponding temperature is given by
\begin{equation}\label{TDS1}
  t_{S} = \frac{5 + 3\sqrt{2}}{2} \Lambda \approx 4.6 \Lambda.
\end{equation}

\subsection{Nonzero magnetic field and phase diagram}
\label{SSmf}

For definiteness we consider only magnetic field along the tetragonal $\m{c}$ axis, which results in finite homogeneous spin component along it. We assume that near $T_c$ the system is far from the ferromagnetic transition critical point, $\Lambda \ll A T_c - \lambda_0$, where $\lambda_0 = (J_\mathbf{0} - \omega_0 \mathcal{N}_{zz})/2$ (we assume ellipsoidal shape of the sample, $\mathcal{N}_{zz}$ being the corresponding demagnetization tensor component~\cite{SpinWaves}). Thus, the spin ordering of each phase acquires correction $\delta \mathbf{s}_i = m \mathbf{e}_z$, which can be determined using the Curie-Weiss law:
\begin{equation}\label{Magn1}
  m = \chi(T) h =  \frac{h}{2(AT - \lambda_0)},
\end{equation}
provided that the high-temperature mean-field expansion~\eqref{Free1} is correct. Note that $\chi(t)$ is almost constant in this region ($T$ close to $T_c$) and can be substituted by $\chi \equiv \chi(T_c)$.

We further proceed with the influence of magnetic field on different magnetic structures. All the relevant spin orderings~\eqref{Ssdw3},~\eqref{Sel1},~\eqref{Sel3} has now additional term $m \mathbf{e}_z$. For the solutions presented above it means appearance of new (proportional to the squared order parameters and squared magnetization) terms originating from $B T_c \sum_i s^4_i$ part of the free energy~\eqref{Free1}. It is easy to show that: (i) for 2S it leads to effective ``temperature'' change $t \rightarrow t^\prime = t - 2 B T_c (\chi h)^2$, (ii) for 1Q and 2Q along with the same substitution ($t \rightarrow t^\prime$) one should also substitute $\Lambda$ with $\Lambda^\prime = \Lambda + 4 B T_c (\chi h)^2$. Importantly, $t^\prime$ and $\Lambda^\prime$ should be directly plugged into free energies~\eqref{Free2}. Another contribution from the magnetic field is identical for all the phases, being equal to $ - \chi h^2/2$, so it can be omitted.

However, one should bear in mind that for conical XY spin ordering at $h=0$ there is no order parameter $z$-component and its interaction with magnetic field leads to different effect. XY structure is similar to 1Q but modulated spin components are in $ab$ plane:
\begin{equation}\label{Scone1}
  \mathbf{s}_i = s_1 \mathbf{e}_y \cos{\mathbf{k}_x \mathbf{R}_i} + s_2 \mathbf{e}_x \sin{\mathbf{k}_x  \mathbf{R}_i} + m \mathbf{e}_z,
\end{equation}
Note, that the spin component $\propto s_2$ is along the hard axis. We denote $\Lambda^{\prime\prime} = \lambda_1 - \lambda_3 > \Lambda$. So the free energy at $h=0$ reads
\begin{eqnarray} \label{Free4}
  f_{XY} &=& -\frac{4 t^2 - 4 \Lambda^{\prime\prime} t + 3 {\Lambda^{\prime\prime}}^2}{16}, \, t>\frac{3 \Lambda^{\prime\prime}}{2}.
\end{eqnarray}
In magnetic field  one should change the ``temperature'' $t$ by $t^\prime$ as for the other relevant phases. However, $\Lambda^{\prime\prime}$ stays intact. This, along with other effects, leads to spiral plane flop (transition 1Q$\leftrightarrow$XY, which is well-known for frustrated antiferromagnets with dipolar interaction, see Ref.~\cite{utesov2018}) at certain $h_{SF}$ for which $\Lambda^\prime=\Lambda^{\prime\prime}$. One obtains
\begin{equation}\label{Hsf}
  h_{SF} = \sqrt{\frac{\Lambda^{\prime\prime} - \Lambda}{4 B T_c \chi^2}},
\end{equation}
which is almost constant upon temperature variation.

Using presented above simple relations, we can derive analytical expressions for the phase boundaries. First, the boundary between PM (or field-induced ferromagnetic-like collinear state) and 2S is given by
\begin{equation}\label{Htc}
  t_c(h) = 2 B T_c \chi^2 h^2.
\end{equation}
Next, second order phase transition curve between 2S and 2Q is the following:
\begin{equation}\label{Hteld}
  t_{2Q}(h) = \frac{5 \Lambda}{2} + 12 B T_c \chi^2 h^2.
\end{equation}
At $h < h_{SF}$ there is also a boundary between 1Q and 2Q phases
\begin{equation}\label{Htds}
  t_{S}(h) = 4.6 \Lambda + 20.4 B T_c \chi^2 h^2.
\end{equation}

Phase boundaries which include XY (see also Eq.~\eqref{Hsf}) are as follows. (i) With the 2S phase it reads
\begin{equation}\label{H2s}
  t_{XY-2S}(h) = 4.1 \Lambda^{\prime\prime} + 2 B T_c \chi^2 h^2.
\end{equation}
(ii) With the 2Q phase the expression is rather cumbersome:
\begin{eqnarray}\label{H2Q}
  t_{XY-2Q}(h) &=& 2 B T_c \chi^2 h^2 \\ &+& \frac{9\Lambda^{\prime\prime} - 4 \Lambda^{\prime} + 3\sqrt{4(\Lambda^{\prime\prime} - \Lambda^{\prime})^2+2{\Lambda^{\prime\prime}}^2}}{2}. \nonumber
\end{eqnarray}
We would like to point out that exact numerical minimization of the free energy~\eqref{Free1} in magnetic field does not change the phase boundaries presented above significantly.

Before considering the phase diagram for particular parameters set, lets have a closer look on Eq.~\eqref{H2Q} at $h=h_{SF}$. In fact, it is determining position of the triple point where 1Q, 2Q and XY are in equilibrium. Using Eqs.~\eqref{H2Q} and~\eqref{Hsf} we get
\begin{equation}\label{Triple}
  t_{tr} \approx (\Lambda^{\prime\prime}-\Lambda)/2 + 4.6 \Lambda^{\prime\prime}.
\end{equation}
The difference between $\lambda_i$ values is usually of the order of $0.1$ K (see Eq.~\eqref{Par1}), which provides an estimation \mbox{$t_{tr} \sim 1$ K} and (using $A \sim 0.1$) $T_c - T_{tr} \sim 10$ K in standard units. In real systems in that region of the phase diagram $|\m{s}_i| \sim S$, and Landau expansion breaks down, thus making predictions involving the conical XY phase unreliable.

\begin{figure}
  \centering
  \includegraphics[width=8cm]{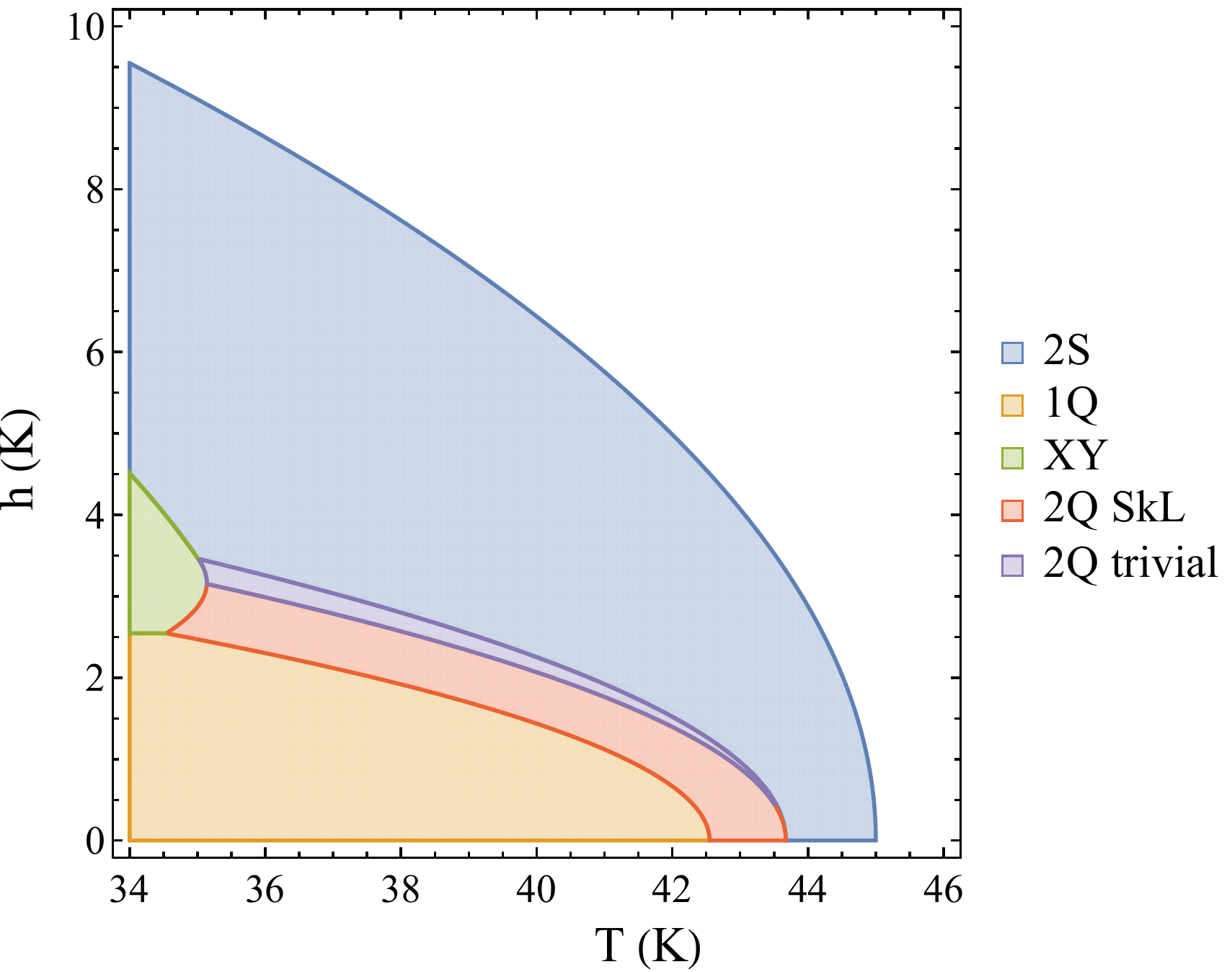}\\
  \caption{High-temperature part of the phase diagram for centrosymmetric tetragonal frustrated antiferromagnet with two possible mutually perpendicular modulation vectors and dipolar interaction (see Fig.~\ref{Fig1}). The parameters~\eqref{Par1} were used. Depending on magnetic field and temperature, the 2Q phase can be either topologically trivial or not (see text). The conical XY phase emerges beyond the theory applicability region and is shown only for illustration purposes.}\label{FPhase1}
\end{figure}

Lets proceed with particular example of the phase diagram for the set of parameters~\eqref{Par1}. In this case \mbox{$\lambda_0 \approx 2.3$ K}, which justifies the approximation of constant susceptibility in the relevant part of the phase diagram, which we draw in Fig.~\ref{FPhase1}. Near the temperature of triple point ($t_{tr} \approx 1.06$ K), where XY can come into play, using Eqs.~\eqref{Sel2} one has $s_1 \approx 2.5$ and $s_2 \approx 2.4$, which means that our approach essentially fails at such temperatures. This rises important question, whether the XY conical phase, which as it is seen in Fig.~\ref{FPhase1} can terminate the 2Q phase region, emerges at low-temperatures in reality or not.

\section{Topological properties of 2Q phase}
\label{Stop}

Using Eq.~\eqref{charge1} it is easy to show that 2S, 1Q, and XY are, as always, topologically trivial; $n_{sk}=0$.

Lets turn to the 2Q structure. First of all using Eq.~\eqref{Sel3} we rewrite the spin ordering in magnetic field in the following form:
\begin{equation}\label{S2Qm1}
  \m{s}(x,y) = \left(
              \begin{array}{c}
                s_1 \sin{ky} \\
                - s_1 \sin{kx} \\
                s_2 [\cos{kx} + \cos{ky}] + m \\
              \end{array}
            \right).
\end{equation}
Magnetic unit cell is a square with the size $(2 \pi /k) \times (2 \pi /k)$. Note, that for illustration purposes we take the one particular structure with $s_1,s_2,m >0$. Its counterparts with other relative phases and chiralities of two elliptical components can be analyzed the similar way. This can be accompanied with change of the signs of the corresponding topological charges (e.g., skyrmion can be substituted by antiskyrmion).

At zero field magnetic ordering has an important antisymmetry property: $\m{s}(x,y) = - \m{s}(x \pm \pi /k, y \pm \pi /k)$. This is equivalent to $ \langle n_{sk} \rangle = 0$ ($\langle ... \rangle$ is averaged over magnetic unit cell quantity). However, the magnetic ordering is somewhat non-trivial, the structure consists of core-down merons with $Q = -1/2$ and core-up antimerons with $Q = + 1/2$ (see Fig.~1 of Ref.~\cite{Yu2018} for the details) alternating in square lattice as it is shown in Fig.~\ref{FigSpins}(d).

\begin{figure}
  \centering
  \includegraphics[width=6cm]{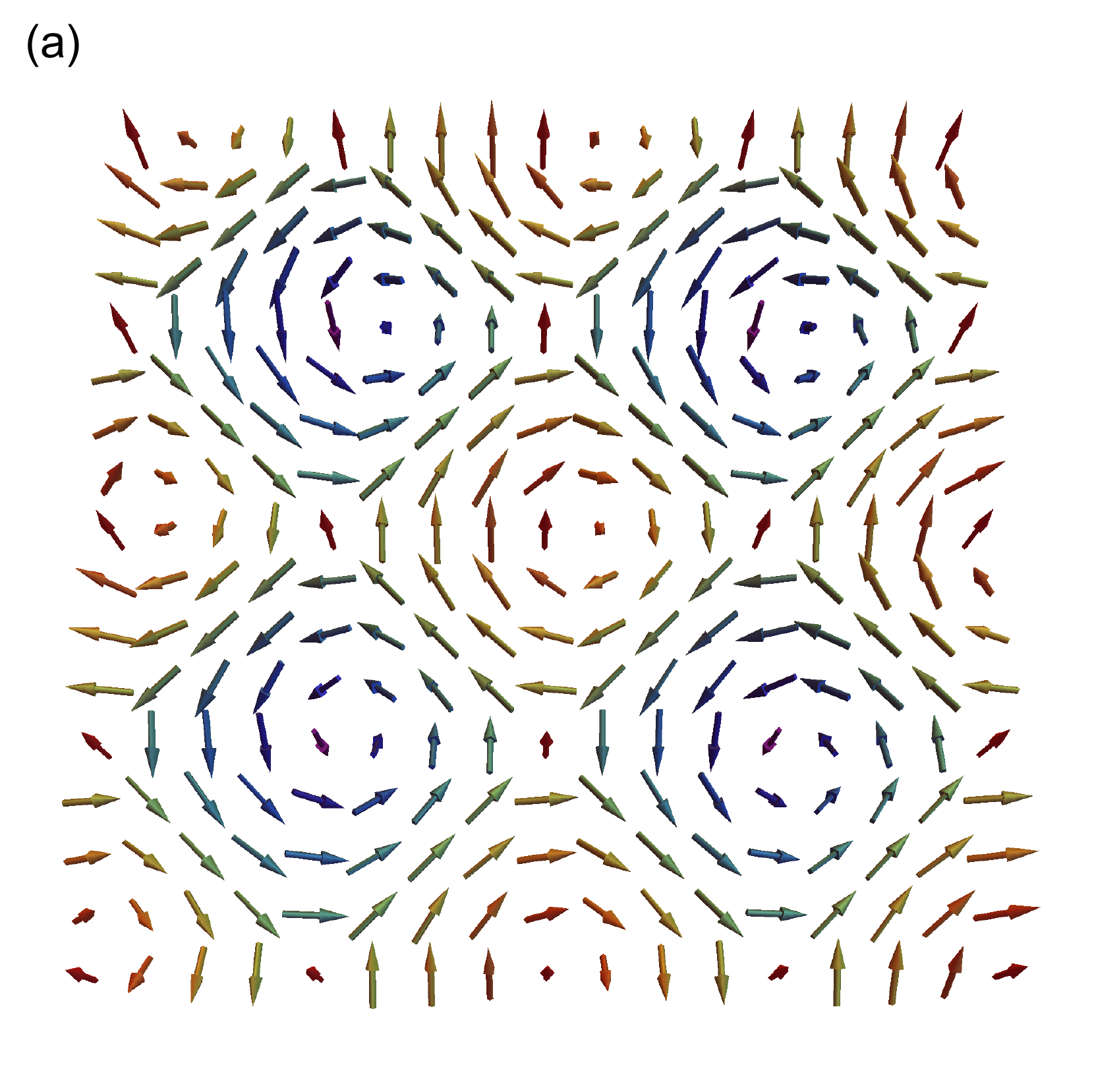}
  \centering
  \includegraphics[width=6cm]{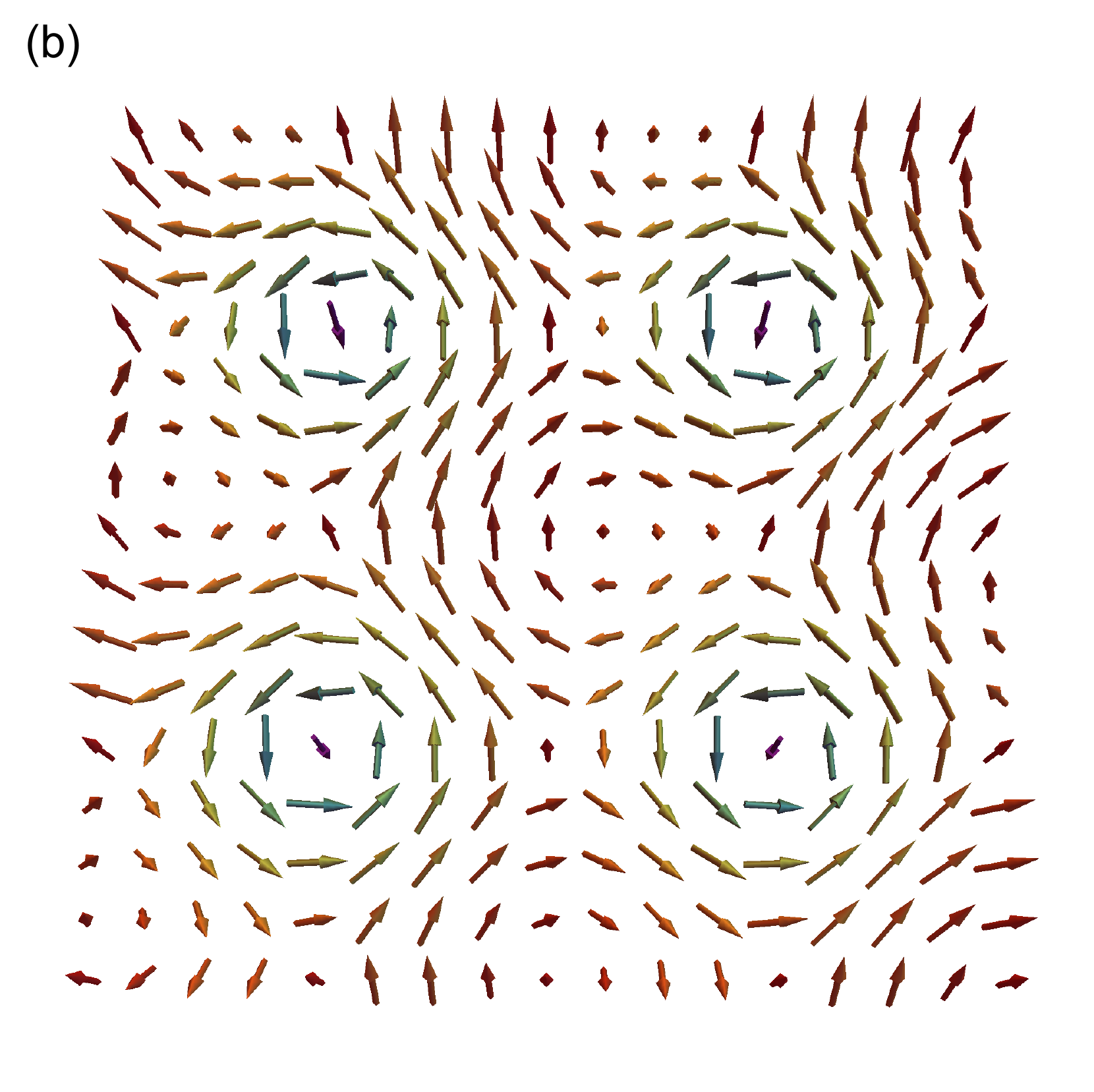}
  \caption{Sketch of the 2Q spin ordering in applied magnetic field; part of the $ab$ plane with size of $2 \times 2$ magnetic unit cells is shown. (a) At small magnetic fields in comparison with zero-field case (cf. Fig.~\ref{FigSpins}(d)) additional small core-up merons emerge providing topological charge $Q=-1$ per unit cell. (b) At larger field boundary between core-up merons and antimerons vanishes, and the magnetic ordering represents square skyrmion lattice. }\label{FigTop}
\end{figure}

Nonzero $h$ brakes the above-mentioned antisymmetry and the magnetic ordering becomes topologically non-trivial with $Q = -1$ per magnetic unit cell. At small $h$ which results in $m \ll s_1, s_2$ the latter can be understood as follows. One can neglect $m$ in spin ordering~\eqref{S2Qm1} almost everywhere except for a small neighborhood (its radius is $\sim \sqrt{m/s_2} \ll 1$) of points with coordinates $(\pi/k, 0)$, $(0,\pi/k)$, and equivalent to them. In these regions core-up merons with $Q = - 1/2$ emerge at $h>0$ (see Fig.~\ref{FigTop}(a)). For the magnetic unit cell one has four halves of such merons, thus $n_{sk} = -1$. At moderate $h$ for which $m \sim s_1, s_2$ the boundary between core-up merons and core-up antimerons is no longer pronounced and the whole magnetic structure can be considered as a square skyrmion lattice, which is shown in Fig.~\ref{FigTop}(b). Under further magnetic field increase, when condition $m < 2 s_2$ violates, the magnetic structure becomes topologically trivial, since all spins $z$ components are positive.

We proceed with the phase diagram established in the previous section. Evidently the whole region of the 2Q phase stability can not be topologically non-trivial because near its boundary with 2S $s_2 \ll 1$. At given $h$, in order to have skyrmion lattice, the condition $4 s^2_2 > m^2 = \chi^2 h^2$ should be fulfilled, where
\begin{equation}\label{S2Qm2}
  s^2_2 = \frac{2t^\prime - 5 \Lambda^\prime}{18 B T_c}.
\end{equation}
Using these formulas one can define the boundary for the SkL region inside the 2Q one as (see Eq.~\eqref{Hteld})
\begin{equation}\label{SkL1}
  t_{SkL}(h) = t_{2Q}(h) + \delta t (h), \quad \delta t(h) = \frac{9}{4} B T_c \chi^2 h^2.
\end{equation}
Importantly it is smaller than $t_S(h)$ (see Eq.~\eqref{Htds}). Fig.~\ref{FPhase1} illustrates these statements.

\section{Mean-field approach for collinear out-of-plane easy axes}
\label{SOOP}

Consideration above relies on small single-ion anisotropy which cannot alter axes hierarchy established by dipolar interaction (see Fig.~\ref{Fig1}(c)). However, it yields substantially different phase diagram shown in Fig.~\ref{FPhase1} in comparison with the experimentally observed one in Ref.~\cite{khanh2020}. Here we consider significant single-ion anisotropy which makes the $\m{c}$ axis an easy one for both modulation vectors $\m{k}_x$ and $\m{k}_y$. Mathematically, in comparison with pure dipolar case ($Z=0$) the eigenvalues change as follows: $\lambda_1 \rightarrow \lambda_1 - Z, \, \lambda_2 \rightarrow \lambda_2 + Z,
\, \lambda_3 \rightarrow \lambda_3 - Z  $. So, for $Z > (\lambda_1 - \lambda_2)/2$ the hard axes stay intact, however, the easy and the middle ones are swapped.

In the mean-field analysis below we continue to use $\lambda_1 \geq \lambda_2 \geq \lambda_3$ bearing in mind that the easy direction is now along the $\m{c}$ axis.

\subsection{Spin structures}

Here we briefly discuss relevant spin structures at both $h=0$ and $h \neq 0$.

\subsubsection{Single-Q spin-density wave (1S)}

The spin ordering of 1S reads
\begin{equation}\label{S1SA1}
  \mathbf{s}_i = s \mathbf{e}_z \cos{\mathbf{k}_x \mathbf{R}_i},
\end{equation}
where
\begin{equation}\label{S1SA2}
  s^2 = \frac{2(\lambda_1-AT)}{3BT_c} = \frac{2 t}{3 B T_c}.
\end{equation}
Corresponding free energy is given by (cf. Subsec.~\ref{SSpth0})
\begin{equation}\label{F1SA1}
  f_{1S} = - \frac{t^2}{6}, \, 3 \Lambda/2 \geq t > 0.
\end{equation}
At larger $t > 3 \Lambda/2$ it transforms into the 1Q structure, see below.

In the external magnetic field one should make the substitution $t \rightarrow t^\prime = t - 6 B T_c (\chi h)^2$.

\subsubsection{Double-Q spin-density wave (2S)}

In comparison with Sec.~\ref{SMF}, here vortical structure involves two middle axes. This immediately affects the phase diagram as it is shown below. The spin structure reads
\begin{equation}\label{S2SA1}
  \mathbf{s}_i = s (\mathbf{e}_x \cos{\mathbf{k}_y \mathbf{R}_i} + \mathbf{e}_y \cos{\mathbf{k}_x \mathbf{R}_i}),
\end{equation}
where
\begin{equation}\label{S2SA2}
  s^2 = \frac{2(\lambda_2-AT)}{5BT_c} = \frac{2(t - \Lambda)}{5BT_c}.
\end{equation}
The free energy is given by
\begin{equation}\label{F2SA1}
  f_{2S} = - \frac{(t - \Lambda)^2}{5}, \, t > \Lambda.
\end{equation}
So, in the considered case there is always range of parameters at which 1S is preferable in comparison with 2S, which should be contrasted to the results of Sec.~\ref{SMF}.

In the magnetic field one should make the change \mbox{$t \rightarrow t^\prime = t - 2 B T_c (\chi h)^2$}.

\subsubsection{Single-Q elliptical phase (1Q)}

In this case spin ordering reads
\begin{equation}\label{S1QA1}
  \mathbf{s}_i = s_1 \mathbf{e}_z \cos{\mathbf{k}_x \mathbf{R}_i} + s_2 \mathbf{e}_y \cos{\mathbf{k}_x \mathbf{R}_i},
\end{equation}
where
\begin{eqnarray}\label{S1QA2}
  s^2_1 &=& \frac{2 (\lambda_1 - A T) + (\lambda_1-\lambda_2)}{4 B T_c}= \frac{2 t + \Lambda}{4 B T_c}, \\ s^2_2 &=& \frac{2 (\lambda_1 - A T) -  3(\lambda_1-\lambda_2)}{4 B T_c} = \frac{2 t -  3\Lambda}{4 B T_c}. \nonumber
\end{eqnarray}
Corresponding free energy is following:
\begin{equation}\label{F1QA1}
  f_{1Q} = - \frac{4t^2 - 4\Lambda t + 3 \Lambda^2}{16}, \, t > 3 \Lambda/2 \wedge t > - \Lambda/2.
\end{equation}
The last inequality becomes important in magnetic field, where one should use \mbox{$t^\prime = t - 6 B T_c (\chi h)^2$} and \mbox{$\Lambda^\prime = \Lambda - 4 B T_c (\chi h)^2$}.

For the XY phase with spins rotating in the $ab$ plane, one has
\begin{eqnarray}  \nonumber
  f_{XY} &=& - \frac{4(t - \Lambda)^2 - 4(t - \Lambda) (\Lambda^{\prime\prime} - \Lambda) + 3  (\Lambda^{\prime\prime} - \Lambda)^2}{16}, \\ && t > \frac{3 \Lambda^{\prime\prime} - \Lambda}{2}. \label{FXYA1}
\end{eqnarray}
In the external field substitution \mbox{$t \rightarrow t^\prime = t - 2 B T_c (\chi h)^2$} should be done, whereas $\Lambda^{\prime\prime}$ and $\Lambda$ stay intact.

\subsubsection{Double-Q elliptical phase (2Q)}

Spin ordering in the double-Q phase is given by
\begin{eqnarray}\label{S2QA1}
  \mathbf{s}_i &=& s_1 \mathbf{e}_z (\cos{\mathbf{k}_x \mathbf{R}_i} + \cos{\mathbf{k}_y  \mathbf{R}_i}) \\
  &&+ s_2 (\mathbf{e}_y \sin{\mathbf{k}_x \mathbf{R}_i} + \mathbf{e}_x \sin{\mathbf{k}_y  \mathbf{R}_i})  \nonumber.
\end{eqnarray}
The order parameters are following:
\begin{eqnarray}\label{S2QA2}
  s^2_1 &=& \frac{2 (\lambda_2 - A T) - 5 (\lambda_2-\lambda_1)}{18 B T_c} = \frac{2 t + 3 \Lambda}{18 B T_c}, \\ \, s^2_2 &=& \frac{2 (\lambda_2 - A T) +  (\lambda_2 - \lambda_1)}{6 B T_c} = \frac{2 t -  3\Lambda}{6 B T_c} \nonumber.
\end{eqnarray}
The free energy has the form:
\begin{equation}\label{F1QA1}
  f_{2Q} = - \frac{8t^2 - 12 \Lambda t + 9 \Lambda^2}{36}, \, t > 3 \Lambda/2 \wedge t > - 3 \Lambda/2.
\end{equation}
As for the 1Q phase, one should use \mbox{$t^\prime = t - 6 B T_c (\chi h)^2$} and \mbox{$\Lambda^\prime = \Lambda - 4 B T_c (\chi h)^2$} in the external magnetic field.

Finally, we note that at $t^\prime = - 3 \Lambda^\prime/2$ (which can be correct only in magnetic field) the 2Q structure continuously transforms into the 2S one.

\subsection{Phase transitions}

In the absence of the external field the sequence of phase transitions is rather trivial in comparison with the one described in Sec.~\ref{SSpth0}. At $t \leq 3 \Lambda /2 $ 1S has lower free energy than 2S. In the complementary domain $t > 3 \Lambda /2 $ 1Q structure free energy is always lower than $f_{2Q}$, which in its turn is lower than $f_{2S}$. So, upon temperature variation at $h=0$ one has \mbox{PM $\leftrightarrow$ 1S $\leftrightarrow$ 1Q} sequence of continuous phase transitions at $t=0$ and $t = 3 \Lambda/2$, respectively.

In the external magnetic field there is following important observation: at $t_0 = 3\Lambda/2$ and $h_0$ for which $ B T_c (\chi h_0)^2 = \Lambda/4$ order parameters of all relevant phases are zero (see previous subsection). The phases PM (equivalently, field polarized phase), 1S, 2S, 1Q, and 2Q are in perfect equilibrium in \emph{polycritical} point; slightly varying $t$ and $h$ one can \emph{continuously} get into each phase.

Now we can derive the phase boundaries at small $t$. First, there is a boundary between PM and 1S at
\begin{equation}\label{PTA1}
  t^{(1)}_c(h) = 6 B T_c \chi^2 h^2, \, h \leq h_0.
\end{equation}
At larger fields the 1S phase does not exist and the PM phase has the boundary with the 2S one:
\begin{equation}\label{PTA2}
  t^{(2)}_c(h) = \Lambda + 2 B T_c \chi^2 h^2, \, h > h_0.
\end{equation}
Next, fixing $h<h_0$ and increasing $t$ one will have continuous phase transition from 1S to 1Q. It is governed by equation $t^\prime = 3 \Lambda^\prime/2$, which for these phases is invariant as a function of $h$, and yields vertical line
\begin{equation}\label{PTA3}
  t_{1Q}(h) = \frac{3 \Lambda}{2}, \, h \leq h_0.
\end{equation}
For $t >t_0$ upon $h$ increase there is a first order transition from 1Q to 2Q at $t^\prime = -3 (1 + \sqrt{2})\Lambda^\prime/2 \approx -3.6 \Lambda^\prime$, or equivalently:
\begin{equation}\label{PTA4}
  t_{S}(h) = -3.6 \Lambda + 20.4 B T_c \chi^2 h^2, \, h> h_0.
\end{equation}
Then, when $t^\prime = - 3 \Lambda^\prime/2$ there is a second order transition between the 2Q and the 2S phases, which yields
\begin{equation}\label{PTA5}
  t_{2Q}(h) = -3 \Lambda/ 2 + 12 B T_c \chi^2 h^2, \, h> h_0.
\end{equation}

The XY phase can emerge in magnetic field via spiral plane flop transition from the 1Q one. It can be shown that the corresponding field is (cf. Eq.~\eqref{Hsf})
\begin{equation}\label{HsfA}
  h_{SF} = \sqrt{\frac{\Lambda^{\prime\prime}}{4 B T_c \chi^2}} > h_0.
\end{equation}
As in Sec.~\ref{SMF} one can estimate the triple point temperature; the counterpart of Eq.~\eqref{Triple} reads
\begin{equation}\label{TripleA}
  t_{tr} \approx  5.1 \Lambda^{\prime\prime} - 3.6 \Lambda,
\end{equation}
which also typically lies out of the theory applicability range (see the discussion in Sec.\ref{SSmf}).

For non-trivial lattice topology in the 2Q phase (see Sec.~\ref{Stop}), the condition $4 s^2_1 > m^2 = \chi^2 h^2$ should hold. Using Eq.~\eqref{S2QA2} we arrive to the same result~\eqref{SkL1} of Sec.~\ref{Stop}, however with different $t_{2Q}(h)$ given by Eq.~\eqref{PTA5}. Importantly, in the present case the condition
\begin{equation}\label{TopA1}
  t_{SkL}(h) = t_{2Q}(h) + 9 B T_c \chi^2 h^2/4 < t_{S}(h)
\end{equation}
provides additional restriction on the topologically nontrivial part of the phase diagram, which approximately reads $ B T_c \chi^2 h^2 > \Lambda/2.9$. So, the square SkL part of the phase diagram starts at certain $t > t_0$, see Fig~\ref{FPhase2}.

\subsection{Qualitative description of $\mathbf{GdRu_2Si_2}$ phase diagram}

\begin{figure}
  \centering
  \includegraphics[width=8cm]{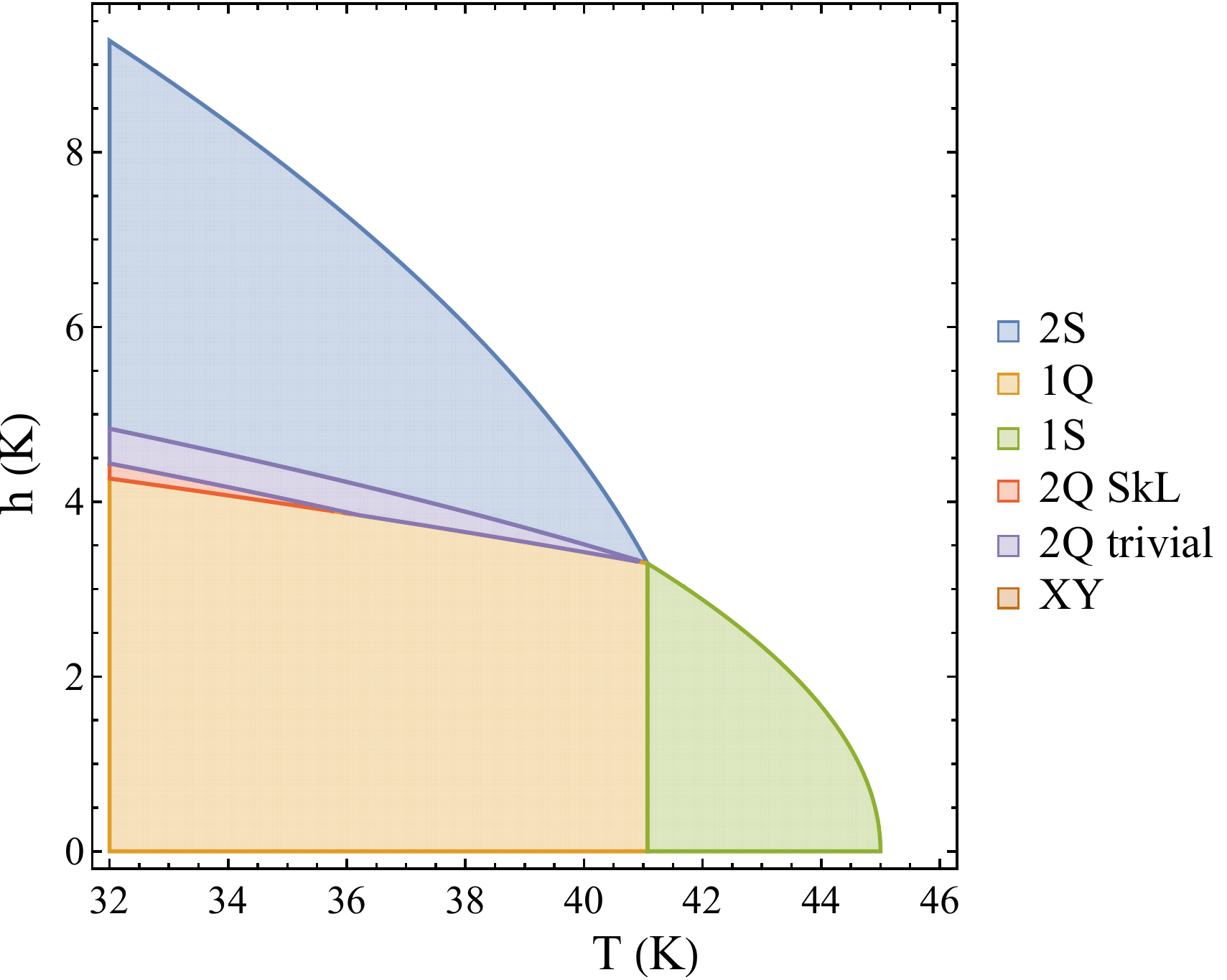}\\
  \caption{Analytically obtained phase diagram for the parameters set~\eqref{Par2}. In comparison with the Fig.~\ref{FPhase1} the easy axes for both modulation vectors are along $\m{c}$ due to the single-ion anisotropy, which leads to crucial differences. In this case the square SkL (red region) emerges only at finite external magnetic field and not very close to the ordering temperature $T_c$. The conical XY phase emerges only at $T \lesssim 30$ K, where the mean-field approach is not applicable. This phase diagram captures important features of the experimentally observed one for GdRu$_2$Si$_2$~\cite{khanh2020}.  }\label{FPhase2}
\end{figure}

Here we utilize parameters of exchange interaction and dipolar tensor from~\eqref{Par1}, however, we add single-ion easy axis anisotropy with $Z = 0.15$ K. It yields (all values are in Kelvins)
\begin{eqnarray} \label{Par2}
  \lambda_1 &\approx& 4.3, \quad \lambda_0 \approx 2.3, \\
  \lambda_1 - \lambda_2 &\approx& 0.25, \quad \lambda_1 - \lambda_3 \approx 0.45,  \nonumber
\end{eqnarray}
where due to the additional anisotropy the easy axis is the $\m{c}$ one.

Obtained phase diagram is shown in Fig.~\ref{FPhase2}. First, we note that in this case the XY phase emerges only at $T \lesssim 30$ K where our approach is inapplicable. Next, the topologically non-trivial square skyrmion lattice is a narrow red wedge in this figure (however, starting at temperatures where the developed approach should work at least qualitatively), which should be contrasted with the large SkL domain for in-plane easy axes, see Fig.~\ref{FPhase1}. Finally, we point out that the phase diagram (Fig.~\ref{FPhase2}) has important similarities with the one of Ref.~\cite{khanh2020}. For instance, its topologically non-trivial narrow part starts at finite magnetic field and at certain temperature not very close to $T_c$. Thus, we suggest that additional experiments determining the phase boundaries in GdRu$_2$Si$_2$ are in order.

\section{Discussion and conclusion}
\label{Sdis}

To conclude, we show that magnetic dipolar interaction can stabilize square skyrmion lattice in centrosymmetric tetragonal frustrated antiferromagnets. The size of the corresponding magnetic unit cell is of the order of several nanometers.

We find that the hierarchy of the axes is crucial for the magnetic field-temperature phase diagram, and provide analytical mean-field consideration of the two possible cases at high temperatures domain. If the easy axes for both modulation vectors are collinear, the phase diagram resembles recently observed one for GdRu$_2$Si$_2$~\cite{khanh2020}. However, there are important analytical predictions which can be checked experimentally: the square SkL region is only a part of the double-Q elliptical phase, which at larger fields continuously transforms into the double-Q vortical structure. Near the latter phase transition the spin component along the external field is always positive and the structure is topologically trivial.

Importantly, in our analysis the conical phase emerges in a certain part of the phase diagram. However, using parameters relevant to GdRu$_2$Si$_2$ we show that our approach fails in that region. Nevertheless, in general, the conical phase can be pronounced in the phase diagram. So, further studies devoted to low-temperatures are important. For example, in Ref.~\cite{utesov2020phase} it was shown that depending on parameters the conical phase can or cannot appear in frustrated antiferromagnets with only single-Q modulated structures possible. Moreover, at small temperatures skyrmion textures contain lots of non-negligible additional harmonics. The construction of the corresponding lattice and its energy calculation, being usually a hard problem itself~\cite{timofeev2019towards}, in the present model with dipolar forces becomes very challenging even numerically due to their long-range character.

\begin{acknowledgments}

We are grateful to V.A.\ Ukleev and A.V.\ Syromyatnikov for valuable discussions. The reported study was supported by the Foundation for the Advancement of Theoretical Physics and Mathematics ``BASIS''.

\end{acknowledgments}

\bibliography{TAFbib}

\end{document}